\documentclass[aps,pra,twocolumn,showpacs,groupedaddress]{revtex4-1}

\usepackage{amsmath} 
\usepackage{graphicx}
\usepackage{amssymb}

\begin{document}

\title{Imbalanced ultracold Fermi gas in the weakly repulsive
  regime: Renormalization group approach for $p$-wave
  superfluidity}

\author{Shao-Jian Jiang}
\author{Xiao-Lu Yu}
\author{W. M. Liu}
\affiliation{Beijing National Laboratory for Condensed Matter Physics,
  Institute of Physics, Chinese Academy of Sciences, Beijing 100190, China}

\date{\today}

\begin{abstract}
We theoretically study a possible new pairing mechanism for a
two-dimensional population imbalanced Fermi gas with short-range
repulsive interactions which can be realized on the upper branch of a
Feshbach resonance. We use a
well-controlled renormalization group approach,
which allows an
unbiased study of the instabilities of imbalanced Fermi liquid without
 assumption of a broken symmetry and  gives
 a numerical calculation of the transition temperature from
microscopic parameters. Our results show a leading
superfluid instability in the $p$-wave
channel for the majority species. The corresponding mechanism is that
there are effective attractive interactions for the majority species,
induced by 
the particle-hole susceptibility of the
minority species, where the mismatch of the Fermi surfaces of the
two species plays an important role.
 We also propose
an experimental protocol for detecting the  $p$-wave superfluidity 
and discuss the
corresponding experimental signatures. 
\end{abstract}


\maketitle

\section{Introduction}

Much of the interest in ultracold atomic gases comes from their
amazing tunability. Experiments on ultracold atomic gases allow
fermionic pairing phenomena to be manipulated much more precisely and 
controllably than those in solid state systems. There are many important 
experiments in ultracold gases which undoubtedly illustrate this
advantage, such as the crossover from Bose-Einstein condensation (BEC)
to Bardeen-Cooper-Schrieffer (BCS) superfluidity with the help of Feshbach
resonance \cite{Regal2004,Zwierlein2004,Chin2004,Kinast2004}, and
superfluid-Mott insulator transitions with optical
lattices \cite{Greiner2002,Stoferle2004}.

Due to the wide-range tunability of the effective interatomic
scattering length there are strong motivations to study the pairing
phenomena with population imbalanced
ultracold Fermi gases in different regimes. However, in ultracold Fermi gases, in contrast with solid state systems, the
pairing state is not easily achieved due to the smallness of the gap
parameter. Therefore, previous investigations on superfluidity of imbalanced Fermi gases
mostly focused on the unitary regime where the
scattering length is large
\cite{Giorgini2008,Zwierlein2006,Gubbels2008,Patton2011}. In systems
with attractive interactions, the
presence of population imbalance can enrich the possibilities for
pairing states. As predicted by previous works, there may be Larkin-Ovchinnikov-Fulde-Ferrell
(LOFF) state \cite{Fulde1964, Larkin1964,James2010}, breached pair
state \cite{Liu2003,Gubankova2003,Forbes2005} and
deformed Fermi surfaces \cite{Muther2002}.
Pairing can also occur when there are intermediate
bosons for providing effective attractive interactions such as bosonic molecules in deep BEC
regime, where there may be $p$-wave superfluidity \cite{Bulgac2006,
  Bulgac2009, Iskin2007}, and phonons of a dipolar condensate
\cite{Kain2011}.
Besides, in a system where the
bare interactions are purely repulsive, there are also possibilities
for effective attractive interactions to emerge. It was first studied
by  Kohn and Luttinger \cite{Kohn1965}, where a
three-dimensional (3D) electron system was 
considered. In 3D electron systems, the particle-hole susceptibility
$\chi(k)$ has a
strong $k$ dependence for $k \leqslant 2k_F$, which is responsible for
the emergence of  effective attractive interactions in high
angular-momentum channel. However, dimensionality can significantly
change the behavior of $\chi (k)$. In two dimension (2D), $\chi(k)$ is momentum
independent when $k \leqslant 2k_F$, and there may be superfluid
instability in the presence of population imbalance \cite{Raghu2011},
which is different from the Kohn-Luttinger type.

In this paper, we consider a population imbalanced 2D ultracold
Fermi gases in the weakly repulsive regime, which can be realized on
the upper branch of a
Feshbach resonance \cite{Jo2009}. 
There are two novelties in our
system that should be emphasized. Firstly, the bare interactions between
atoms of two different hyperfine states
are purely repulsive. Secondly, there are no intermediate
bosons for providing effective attractive interactions such as phonons
in traditional superconductors or bosonic molecules at the BEC side of
a Feshbach resonance.
Our study shows that there is an alternative choice
of $p$-wave superfluid state induced by the population imbalance, which
fundamentally breaks the spin rotation symmetry. The particle-hole
susceptibility of the minority species can induce an attractive
interaction for the majority species because of the population imbalance.
This mechanism of
superfluidity resembles qualitatively the situation  in the $A_1$ phase
of superfluid $^{3}\mathrm{He}$ \cite{Vollhardt1990} and 2D
electronic gases \cite{Raghu2010,Raghu2011}.

Our theoretical framework is heavily based on the
renormalization group (RG) theory for interacting fermion
systems \cite{Shankar1991,Shankar1994,Polchinski1994,Weinberg1994}. The
RG framework provides us a powerful tool to treat competing
instabilities simultaneously, and most importantly, to justify the
 leading instability channel \cite{Shankar1994}. Furthermore, we can
identify the critical temperature from the onset of the instability
channel \cite{Tsai2005}. By performing RG process at finite temperature
 and solving the flow equations numerically, we can obtain the
phase transition  between normal state and $p$-wave superfluid
state. Within this framework, in the second stage of RG, when mode
eliminations have reached an momentum cutoff $\Lambda$ far smaller than
the Fermi momentum $ K_{F}$, a large-N expansion emerges with
$N=K_{F}/\Lambda$, which is a strong suggestion for us to extend our
results from weak to intermediate coupling
regime \cite{Shankar1994}.

The paper is organized as follows: The first stage of the RG approach for
 building the model of interacting imbalanced fermions is described in
Sec. II. Sec. III illustrates the non-perturbative RG method for
unequal Fermi surfaces and obtain the flow
equations. The RG analysis indicates a leading instability in the
$p$-wave Cooper channel when the population imbalance is present. In
Sec. IV we numerically solve the flow equation at finite
temperature. We obtain the critical temperature at which the normal
Fermi liquid state becomes unstable in the $p$-wave Cooper
channel. Furthermore, with the large-$N$ analysis, we extend our results
from weak coupling regime to intermediate coupling regime where we may
have higher critical temperature. Sec. V
contains experimental discussions and conclusions.

\section{Model building: The first stage of RG}

We consider a population imbalanced Fermi gas with short-range
Hubbard repulsive interactions, whose partition function
\begin{equation}
  \label{eq:30}
  \mathcal{Z}=\int\,D[\bar{\psi},\psi] e^{-S[\bar{\psi},\psi]}
\end{equation}
with
\begin{equation}
  \label{eq:1}
  S[\bar{\psi},\psi]=S_{0}[\bar{\psi},\psi]+S_{I}[\bar{\psi},\psi],
\end{equation}
where $S_0$ is the free part,
\begin{equation}
  \label{eq:2}
  S_{0}[\bar{\psi},\psi]=\sum_{k,\sigma}\bar{\psi}_{\sigma}(k)
  (-ik_n-\mu_{\sigma}+E(\boldsymbol{k})) \psi_{\sigma}(k).
\end{equation}
 $k$ is short for $(k_n,\boldsymbol{k})$.
$\sigma=\uparrow$ or $\downarrow$ represents two different hyperfine states. 
$E(\boldsymbol{k})=\boldsymbol{k}^2 /(2m)$ is the free energy of
atoms. $\mu_{\sigma}$ is the chemical potential, and the Fermi
momentum satisfies $\mu_{\sigma}=K_{F\sigma}^2/(2m)$. 
Population imbalance is put in by setting $\mu_{\uparrow} \neq
\mu_{\downarrow}$ ( Without loss of generality, we can assume
$\mu_{\uparrow}>\mu_{\downarrow}$. ).
We work at 
finite temperature, with imaginary time formulism, where $k_n$ is the
fermionic Matsubara frequency. The interacting part of the action reads
\begin{eqnarray}
  \label{eq:3}
  S_{I}[\bar{\psi},\psi]&=&\frac{U}{\beta V}  \sum_{\{k_i\}}
  \bar{\psi}_{\uparrow}(k_1)
  \bar{\psi}_{\downarrow}(k_2)
  \psi_{\downarrow}(k_3)
  \psi_{\uparrow}(k_4) \nonumber \\
  && \times \delta_{k_{1n}+k_{2n},k_{3n}+k_{4n}}
  \delta(\boldsymbol{k}_1+\boldsymbol{k}_2-\boldsymbol{k}_3-
  \boldsymbol{k}_4).
\end{eqnarray}

The basic idea of RG is to gradually integrate out ``faster'' degrees
of freedom which have larger momentums locating in a shell region in
momentum space and see how the resulting effective Hamiltonian will
flow under such process.  To begin the RG process, we first introduce an
artificial 
energy scale $\Omega_0$ and integrate out degrees of freedom with
energies higher than $\Omega_0$ to obtain the effective theory of the
system at energy scale $\Omega_0$.  We assume that the bare
Hamiltonian is in the weak coupling regime. Thus, if we choose
$\Omega_0$ not much lower than the ultraviolet cutoff of the bare
Hamiltonian, the ``integrating out'' procedure can be done by using
straightforward perturbative approach because there is no significant
renormalization of the 
coupling parameters.

In detail, we first divide the degrees of freedom of the system into
``slow modes'' 
\begin{equation}
\label{eq:38}
 \psi_{\sigma}^{<} = \psi_{\sigma}(k),\bar{\psi}_{\sigma}^{<} =
  \bar{\psi}_{\sigma}(k) \;
  \text{for} \; |\varepsilon_{\boldsymbol{k},\sigma}|< \Omega_0
\end{equation}
and ``fast modes'' 
\begin{equation}
\label{eq:39}
\psi_{\sigma}^{>} = \psi_{\sigma}(k), \bar{\psi}_{\sigma}^{>} =
\bar{\psi}_{\sigma}(k) \;  \text{for} \; |\varepsilon_{\boldsymbol{k},\sigma}|>
\Omega_0,
\end{equation}
where $\varepsilon_{\boldsymbol{k},\sigma}=E(\boldsymbol{k})-\mu_{\sigma}$.

We then carry out the ``modes elimination'' by integrating out the
fast modes and this can be formally written as
\begin{eqnarray}
  \label{eq:7}
  Z & = & \int D[\bar{\psi},\psi] \mathrm{e}^{-S[\bar{\psi},\psi]} \nonumber \\
  & = & \int D[\bar{\psi}^{<},\psi^{<},\bar{\psi}^{>},\psi^{>}]
  \nonumber \\
  && \times \; \mathrm{e}^{-S[\bar{\psi}^{<} ,\psi^{<}]-S[\bar{\psi}^{>} ,\psi^{>}]-S_2[\bar{\psi}^{<} ,\psi^{<},\bar{\psi}^{>} ,\psi^{>}]} \nonumber \\
  & = & \int D[\bar{\psi}^{<},\psi^{<}] \mathrm{e}^{-S[\bar{\psi}^{<}
    ,\psi^{<}]} \nonumber \\
  && \times \; \int D[\bar{\psi}^{>},\psi^{>}] \mathrm{e}^{-S[\bar{\psi}^{>}
    ,\psi^{>}]-S_2[\bar{\psi}^{<} ,\psi^{<},\bar{\psi}^{>}
    ,\psi^{>}]} \nonumber \\
  & \equiv & \int D[\bar{\psi}^{<},\psi^{<}] \mathrm{e}^{-S^{\Omega_0}[\bar{\psi}^{<} , \psi^{<}]}.
\end{eqnarray}
After gathering all terms independent of $\bar{\psi}^{>}$ and
$\psi^{>}$ into $S[\bar{\psi}^{<},\psi^{<}]$ , the rest terms can be
written as $-S[\bar{\psi}^{>} ,\psi^{>}]-S_2[\bar{\psi}^{<}
,\psi^{<},\bar{\psi}^{>} ,\psi^{>}]$, and  $S^{\Omega_0}$ is the
resulting effective action at energy scale 
$\Omega_0$.

Generally, $S^{\Omega_0}$ has the form:
\begin{equation}
  \label{eq:11}
  S^{\Omega_0}[\bar{\psi},\psi]=S_{0}^{\Omega_0}[\bar{\psi},\psi]+S_{I}^{\Omega_0}[\bar{\psi},\psi],
\end{equation}
where $S_{0}^{\Omega_0}[\bar{\psi},\psi]$ is the free action,
\begin{equation}
  \label{eq:12}
  S_{0}^{\Omega_0}[\bar{\psi},\psi]=\sum_k^{\Omega_0}
  \,\bar{\psi}(k)(-ik_n-\mu+\tilde{E}(\boldsymbol{k}))\psi(k)
\end{equation}
and $S_{I}^{\Omega_0}[\bar{\psi},\psi]$ is the interacting action which
will have the most generic form after the ``integrating out'' procedure.
\begin{eqnarray}
  \label{eq:13}
  S_{I}^{\Omega_0}[\bar{\psi},\psi]&=&\frac{1}{\beta V}
    \sum_{\{k_i,\sigma_i\}}^{\Omega_0}u(k_1,\sigma_1,k_2,
  \sigma_2,k_3,\sigma_3,k_4,\sigma_4)
  \nonumber \\
  &&
  \times \bar{\psi}_{\sigma_1}(k_1) \bar{\psi}_{\sigma_2}(k_2)
  \psi_{\sigma_3}(k_3) \psi_{\sigma_4}(k_4) \nonumber \\
  && \times \delta_{k_{1n}+k_{2n},k_{3n}+k_{4n}}
  \delta_{\sigma_1+\sigma_2-\sigma_3-\sigma_4,0}
  \nonumber \\
  && \times \delta(\boldsymbol{k}_1+\boldsymbol{k}_2-\boldsymbol{k}_3-
  \boldsymbol{k}_4).
\end{eqnarray}
where the superscript $\Omega_0$ of the summation operator means the
summation is done within the slow modes characterized by the energy
scale $\Omega_0$ (Eq. \eqref{eq:38}).

\begin{figure}[t]
  \includegraphics[width=3in]{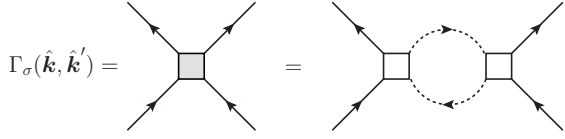}
  \caption{\label{fig:induce} Interaction vertex within the
    same spin species at order $U^2$. It is induced by the
    susceptibility of ``fast modes'' with opposite spins. The
    propagators of fermions with spins opposite 
    to the external ones are represented by
    dashed lines.}
\end{figure}

With singlet
pairing suppressed by imbalance, we shall consider triplet pairing
between fermions with the same spin. As shown in
Fig.~\ref{fig:induce}, the 
induced interaction  within the same spin species in the Cooper
channel is of order $U^2$
and  depends
on  external momentums and frequencies. However, only the constant
term of a coupling function is not 
irrelevant in the tree level scaling \cite{Shankar1994}. Therefore, we
will focus on
the induced Cooper channel effective interaction of order $U^2$ with
external momentums set on the Fermi surface and external frequencies
set to
zero. In this case, the interaction vertex will only depend on the
orientations of 
the incoming and outgoing momentums and can be written as
\begin{equation}
  \label{eq:4}
  \Gamma_{\sigma}( \hat{\boldsymbol{k}} , \hat{\boldsymbol{k}}^{\prime}
  ) = U^2\chi_{-\sigma}(\boldsymbol{k} - \boldsymbol{k}^{\prime}).
\end{equation}

$\chi_{\sigma}(\boldsymbol{k})$ is the susceptibility at zero
frequency:
\begin{eqnarray}
  \label{eq:9}
  \chi_{\sigma}(\boldsymbol{k} ) & = & \int_p
  G_{\sigma}(ip_n , \boldsymbol{p}) G_{\sigma}(ip_n ,
  \boldsymbol{p+k})\nonumber \\
  & = & \int \frac{\mathrm{d}^2 p }{(2 \pi)^2} \frac{f
    (\varepsilon_{\boldsymbol{p+k},\sigma}) -
    f(\varepsilon_{\boldsymbol{p},\sigma})}{\varepsilon_{\boldsymbol{p+k},\sigma}
    - \varepsilon_{\boldsymbol{p},\sigma}},
\end{eqnarray}
where $f$ is the Fermi-Dirac distribution function, and
$\int_p\equiv\frac{1}{\beta}\sum_{p_n} \int \frac{\mathrm{d}^2 p }{(2
  \pi)^2}$. Because the susceptibility is not singular in the limit
$\Omega_0 \rightarrow 0$ ($\Omega_0\ll\mu $), we can only keep the
zeroth order term by setting $\Omega_0=0$. 

\section{The second stage of RG}

\begin{figure}[t]
  \includegraphics[scale=.5]{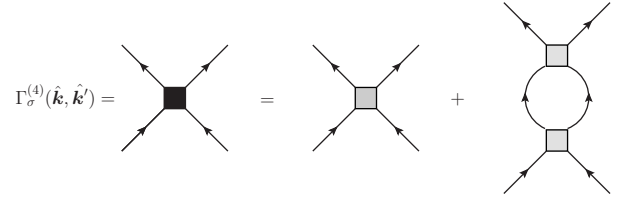}
  \caption{\label{fig:rg} Four-point vertex for the effective action
    at energy scale $\Omega_0$ in the Cooper channel at
    one-loop order. Loop momentums are restricted in the thin shell
    around Fermi surface.}
\end{figure}

After obtaining the effective action at energy scale $\Omega_0$ around
the Fermi surface, 
Shankar's RG \cite{Shankar1994} for fermions can be carried out. We
use the field-theory approach and calculate the four-point vertex at
one-loop order at energy scale $\Omega$ (see Fig.~\ref{fig:rg} ):
\begin{eqnarray}
  \label{eq:10}
  \Gamma^{(4)}_{\sigma}(\hat{\boldsymbol{k}}, \hat{\boldsymbol{k}^{\prime}})
  & = & \Gamma_{\sigma}( \hat{\boldsymbol{k}} ,
  \hat{\boldsymbol{k}^{\prime}}) \nonumber \\
  && - \int_{p}^{\Omega} \frac{\Gamma_{\sigma}(
    \hat{\boldsymbol{k}} , \hat{\boldsymbol{p}} ) \Gamma_{\sigma}(
    \hat{\boldsymbol{p}} , \hat{\boldsymbol{k}^{\prime}} )}
  {(ip_n-\varepsilon_{\boldsymbol{p}, \sigma})
    (-ip_n-\varepsilon_{\boldsymbol{p}, \sigma})},
\end{eqnarray}
where $\int_p^{\Omega}$ means the momentum integral is restricted in
the shell region in momentum space with energy deviation less than
$\Omega$ with 
respect to Fermi energy, and $\Omega$ is a energy scale within
$\Omega_0$, i.e. $\Omega<\Omega_0$. The one-loop correction can be
further carried out as:
\begin{eqnarray}
  \label{eq:15}
  && -\frac{1}{\beta} \sum_n
  \int^{\Lambda_{\sigma}}
  \frac{p\mathrm{d}p}{2\pi} \int_0^{2\pi} \frac{\mathrm{d}\theta}{2
    \pi}
  \frac{\Gamma_{\sigma}(
    \hat{\boldsymbol{k}} , \hat{\boldsymbol{p}} ) \Gamma_{\sigma}(
    \hat{\boldsymbol{p}} , \hat{\boldsymbol{k}^{\prime}} )}
  {(ip_n-\varepsilon_{\boldsymbol{p}, \sigma})
    (-ip_n-\varepsilon_{\boldsymbol{p}, \sigma})}   \nonumber\\
  & = & - \! \int^{\Lambda_{\sigma}} \!
  \frac{p\mathrm{d}p}{2\pi} \int_0^{2\pi} \frac{\mathrm{d}\theta}{2
    \pi}
  \frac{\Gamma_{\sigma}(
    \hat{\boldsymbol{k}} , \hat{\boldsymbol{p}} ) \Gamma_{\sigma}(
    \hat{\boldsymbol{p}} , \hat{\boldsymbol{k}^{\prime}} )}
  {2 \varepsilon_{\boldsymbol{p}\sigma}} \tanh \frac{\beta
    \varepsilon_{\boldsymbol{p}\sigma}}{2}
\end{eqnarray}
where $\Lambda_{\sigma}$ is the momentum cutoff corresponding to
$\Omega$,
and $\int^{\Lambda_{\sigma}}$ is short for
$\int_{|p-K_{F\sigma}|<\Lambda_{\sigma}}$.
Due to $\Omega \ll \mu_{\uparrow(\downarrow)}$, we can approximate
$\Omega$ as $v_{F\sigma} \Lambda_{\sigma}$, where $v_{F\sigma}$ is the
Fermi velocity of atoms with spin $\sigma$.

Because the four-point vertex is related to the scattering amplitude of
certain scattering process, which is a physical observable, it should
not depend on cutoff:
\begin{equation}
  \label{eq:14}
  \frac{\mathrm{d}\Gamma^{(4)}_{\sigma}(\hat{\boldsymbol{k}},
    \hat{\boldsymbol{k}^{\prime}})}
  {\mathrm{d} l}=0,
\end{equation}
where $l=\ln(\Omega_0/\Omega)=\ln(\Lambda_0/\Lambda)$.
 
From Eq.~\eqref{eq:14}, we can get the flow equation as following
\begin{equation}
  \label{eq:16}
  \frac{\mathrm{d} \Gamma_{\sigma}( \hat{\boldsymbol{k}} ,
    \hat{\boldsymbol{k}^{\prime}})} {\mathrm{d} l} = 
  -\rho \! \int_0^{2 \pi} \! \frac{\mathrm{d} \theta}{2 \pi}
  \Gamma_{\sigma}(
  \hat{\boldsymbol{k}} , \hat{\boldsymbol{p}} ) \Gamma_{\sigma}(
  \hat{\boldsymbol{p}} , \hat{\boldsymbol{k}^{\prime}} ) \tanh
  \frac{\beta \Omega}{2} 
\end{equation}
where $\rho=m/2 \pi$ is the 2D density of states.  This is
an ordinary differential equation for matrix $\Gamma_{\sigma}(
\hat{\boldsymbol{k}} , \hat{\boldsymbol{k}^{\prime}})$ with initial
condition:
\begin{equation}
  \label{eq:5}
  \Gamma_{\sigma}( \hat{\boldsymbol{k}} ,
  \hat{\boldsymbol{k}^{\prime}}; \Omega_0) =
  U^2\chi_{-\sigma}(\boldsymbol{k} - \boldsymbol{k}^{\prime}),
\end{equation}
where $\Gamma_{\sigma}( \hat{\boldsymbol{k}} ,
\hat{\boldsymbol{k}^{\prime}}; \Omega_0)$ denotes the effective
interaction vertex at energy scale $\Omega_0$. For convenience, we 
define a dimensionless coupling function
$g_{\sigma}(\hat{\boldsymbol{k}} , \hat{\boldsymbol{k}^{\prime}};
\Omega_0) \equiv \rho \Gamma_{\sigma}( \hat{\boldsymbol{k}} ,
\hat{\boldsymbol{k}^{\prime}}; \Omega_0)$.  In the presence of
rotational symmetry, $g_{\sigma}(
\hat{\boldsymbol{k}} , \hat{\boldsymbol{k}^{\prime}})$  only
depends on the relative angle between the incoming and outgoing
momentum, and the flow equation can be decomposed into uncoupled
equations for eigenvalues of channels with different angular momentums
\cite{Weinberg1994}:
\begin{equation}
  \label{eq:22}
  \frac{\mathrm{d} \lambda_{\sigma m}} {\mathrm{d} l} =
  -(\lambda_{\sigma m})^2 \tanh \frac{\beta \Omega}{2},
\end{equation}
where $m$ labels different angular momentum channels.

The right hand side of Eq.~\eqref{eq:22} is negative definite, which means
that for an initially attractive channel, $\lambda_{\sigma m}$ may be
renormalized to negative infinity as the energy scale goes down to the Fermi
energy. A qualitative argument of the critical
temperature can be given based on Eq.~\eqref{eq:22}.
 At low temperatures, $\tanh(\beta \Omega /2)$ equals almost
unity for nearly all $\Omega$'s when $\Omega>0$, and drops rapidly to zero as $\Omega$
approaches zero from about $\Omega \sim k_BT$. Therefore, when temperature is low
enough, we can approximate $\tanh(\beta \Omega /2)$ to be unity, and
easily get the solution of Eq.~\eqref{eq:22}, which guarantees a
divergence at a certain energy scale. If this energy scale is higher
than $k_BT$, the approximation used above is self-consistent, and the
divergence  indicates the
superfluid instability. In other words, this energy scale gives a
qualitative estimation of the critical temperature.


Results obtained by using the methods introduced above will certainly
depend on the 
energy scale $\Omega_0$. However, it is clear that $\Omega_0$ is more a
calculation device than a physical energy scale \cite{Raghu2010}. Any physical
predictions should not depend on $\Omega_0$. In fact, like RG in
quantum field theory, we can make results independent of $\Omega_0$ at
any order of $U$. This can be achieved  by simply considering diagrams of the same order as
the ones in the second stage when carrying out the perturbative
calculations in the first stage \cite{Raghu2010,Efremov2000}.

\begin{figure}[t]
  \centering
  \includegraphics[width=3in]{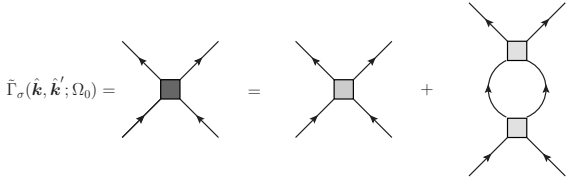}
  \caption{ Induced interaction vertex at energy scale $\Omega_0$
    after reconsidering a $U^4$ 
    order correction, which is responsible for removing the $\Omega_0$
    dependence. The loop modes are the ``fast'' ones, whose
    corresponding on-shell energies are larger than $\Omega_0$.}
  \label{fig:ome}
\end{figure}

In detail, we have to take an additional term into account
(see Fig.~\ref{fig:ome}) and take $\tilde{\Gamma}_{\sigma}(
\hat{\boldsymbol{k}} , \hat{\boldsymbol{k}^{\prime}} ;\Omega_0)$ as
the initial condition for the second stage.

\begin{eqnarray}
  \label{eq:23}
  && \tilde{\Gamma}_{\sigma}( \hat{\boldsymbol{k}} ,
  \hat{\boldsymbol{k}^{\prime}};\Omega_0) \nonumber \\
  & = & \Gamma_{\sigma}( \hat{\boldsymbol{k}} ,
  \hat{\boldsymbol{k}^{\prime}}) -
  \int^{>\Omega_0}_{p} \frac{\Gamma_{\sigma}(
    \hat{\boldsymbol{k}} , p)
    \Gamma_{\sigma}( p ,
    \hat{\boldsymbol{k}^{\prime}})}
  {(ip_n-\varepsilon_{\boldsymbol{p}, \sigma})
    (-ip_n-\varepsilon_{\boldsymbol{p}, \sigma})},
\end{eqnarray}
which has a similar form as Eq.~\eqref{eq:10} except that the region of
the momentum integral
 is the area besides the thin shell around the Fermi
surface and the dependence of the vertex on the
magnitudes of momentums and frequencies should also be considered.

We first consider the frequency summation in the second term of
Eq.~\eqref{eq:23}. Most contribution comes from neighborhood of
$p_0\sim 0$ and $|\boldsymbol{p}|\sim K_{F\sigma}$, so we can first
set frequencies in $\Gamma's$ to zero. Then the frequency summation can be
carried out as:
\begin{eqnarray}
  \label{eq:33}
  && \tilde{\Gamma}_{\sigma}( \hat{\boldsymbol{k}} ,
  \hat{\boldsymbol{k}^{\prime}};\Omega_0) \nonumber \\
  & = & \Gamma_{\sigma}( \hat{\boldsymbol{k}} ,
  \hat{\boldsymbol{k}^{\prime}}) -
  \int_{\Lambda_{\sigma0}}
  \frac{p\mathrm{d}p}{2\pi} \int_0^{2\pi} \frac{\mathrm{d}\theta}{2
    \pi}
  \frac{\Gamma_{\sigma}(
    \hat{\boldsymbol{k}} , \boldsymbol{p} ) \Gamma_{\sigma}(
    \boldsymbol{p} , \hat{\boldsymbol{k}^{\prime}} )}
  {2 \varepsilon_{\boldsymbol{p}\sigma}} \nonumber \\
  && \times \tanh \frac{\beta \varepsilon_{\boldsymbol{p}\sigma}}{2},
\end{eqnarray}
where $\Lambda_{\sigma 0}=\Omega_0/v_{F \sigma}$.

Because the bare interaction is short-range and the Fermi
gas is dilute which means the interatomic distance is much larger than
the range of the interatomic interaction i.e. $K_{F\uparrow}^{-1}\gg
r_0$, we can introduce an ultraviolet cutoff
$\Lambda_H=2K_{F\uparrow}$ here. In the second term of
Eq.~\eqref{eq:33} $\Gamma_{\uparrow}$ does
not vary much with respect to momentum within this cutoff \cite{Raghu2011} at low
temperatures compared with the notable dependence of the other
term on momentum. Therefore, we can neglect the dependence of the
$\Gamma's$ on 
the magnitude of the momentums. Using the dimensionless coupling
function which has been defined above as
$g_{\sigma}(\hat{\boldsymbol{k}} , \hat{\boldsymbol{k}^{\prime}};
\Omega_0) \equiv \rho \Gamma_{\sigma}( \hat{\boldsymbol{k}} ,
\hat{\boldsymbol{k}^{\prime}}; \Omega_0)$ the initial condition of
the flow equation can be written as
\begin{eqnarray}
  \label{eq:26}
  && \tilde{g}_{\sigma}( \hat{\boldsymbol{k}} ,
  \hat{\boldsymbol{k}^{\prime}};\Omega_0) \nonumber \\
  & = & g_{\sigma}( \hat{\boldsymbol{k}} ,
  \hat{\boldsymbol{k}^{\prime}}) -
  \int_0^{2 \pi}\frac{\mathrm{d} \theta}{2 \pi}
  g_{\sigma}(
  \hat{\boldsymbol{k}} , \hat{\boldsymbol{p}} ) g_{\sigma}(
  \hat{\boldsymbol{p}} , \hat{\boldsymbol{k}^{\prime}} ) \nonumber \\
  && \times
  \left( F_{\sigma}(\Omega_H) - F_{\sigma}(\Omega_0) \right),
\end{eqnarray}
where $F_{\sigma}(\Omega)$ is an auxiliary function defined as
\begin{equation}
  \label{eq:24}
  F_{\sigma}(\Omega)=\frac{1}{\rho}\int_{|\varepsilon_{\boldsymbol{p} \sigma}|<\Omega}\frac{p
    \mathrm{d}p}{2 \pi}\frac{1}{2
    \varepsilon_{\boldsymbol{p} \sigma}} \tanh \frac{\beta \varepsilon_{\boldsymbol{p}\sigma}}{2}
\end{equation}

With the help of Eq.~\eqref{eq:24}, flow equation can be expressed as:
\begin{eqnarray}
  \label{eq:25}
  \frac{\mathrm{d} g_{\sigma}( \hat{\boldsymbol{k}} ,
    \hat{\boldsymbol{k}^{\prime}})} {\mathrm{d} \Omega} & =  &
  \int_0^{2 \pi}\frac{\mathrm{d} \theta}{2 \pi}
  g_{\sigma}(
  \hat{\boldsymbol{k}} , \hat{\boldsymbol{p}} ) g_{\sigma}(
  \hat{\boldsymbol{p}} , \hat{\boldsymbol{k}^{\prime}} )
  \frac{\mathrm{d} F_{\sigma} (\Omega)}{\mathrm{d} \Omega}.
\end{eqnarray}

It can also be written in a more compact form, regarding $g_{\sigma}(
\hat{\boldsymbol{k}} , \hat{\boldsymbol{k}^{\prime}})$ as a matrix
$(\mathbf{g}_{\sigma})_{\hat{\boldsymbol{k}} , \hat{\boldsymbol{k}^{\prime}}}$
\cite{Weinberg1994},
\begin{equation}
  \label{eq:17}
  \frac{\mathrm{d} \mathbf{g}_{\sigma}}{\mathrm{d} \Omega} = 
  \mathbf{g}_{\sigma} \cdot \mathbf{g}_{\sigma}
  \frac{\mathrm{d} F_{\sigma} (\Omega)}{\mathrm{d} \Omega}.
\end{equation}

As  we mentioned earlier, because of the rotational symmetry of the
system, the coupling
function can be decoupled in the angular momentum representation, and
Eq.~\eqref{eq:17} becomes a series of flow equations of individual
eigenvalues, 
\begin{equation}
  \label{eq:27}
  \frac{\mathrm{d} \lambda_{\sigma m}}{d \Omega} = 
  (\lambda_{\sigma m})^2 \frac{\mathrm{d} F_{\sigma}}{\mathrm{d} \Omega}
\end{equation}
with innitial conditions,
\begin{equation}
\label{eq:37}
  \lambda_{\sigma m}(\Omega_0)  =  \lambda_{\sigma m} -
  (\lambda_{\sigma m})^2 \left( F_{\sigma}(\Omega_H) -
    F_{\sigma}(\Omega_0) \right),
\end{equation}
which can be easily integrated and gives
\begin{eqnarray}
  \label{eq:28}
  \lambda_{\sigma m}(\Omega)^{-1} & = &
  \left(
    \lambda_{\sigma m}-(\lambda_{\sigma m})^2(F_{\sigma}(\Omega_H)-F(\Omega_0))
  \right)^{-1}  \nonumber \\
  &&  -F(\Omega)+F(\Omega_0).
\end{eqnarray}
To order $U^4$, we have
\begin{equation}
  \label{eq:29}
  \lambda_{\sigma m}(\Omega)=(\lambda_{\sigma
    m}^{-1}+F_{\sigma}(\Omega_H)-F_{\sigma}(\Omega))^{-1},
\end{equation}
which is independent of $\Omega_0$, and controls the flow behavior of
the coupling strengths. At zero temperature, as mentioned above, a
negative  eigenvalue
will flow to infinity at certain energy scale and cause superfluid
instability. As temperature goes higher, the divergence will appear at
lower energy scale. When temperature reaches a certain critical value,
the divergence will not 
arise until we renormalize to the Fermi surface. Above this critical 
value, no divergence exists during the whole 
renormalization process down to the Fermi surface, which means that
the Fermi liquid state is stable in the corresponding channel.

\section{Numerical solutions of RG flow equations}

\begin{figure}[t]
  \centering
  \includegraphics[width=3in]{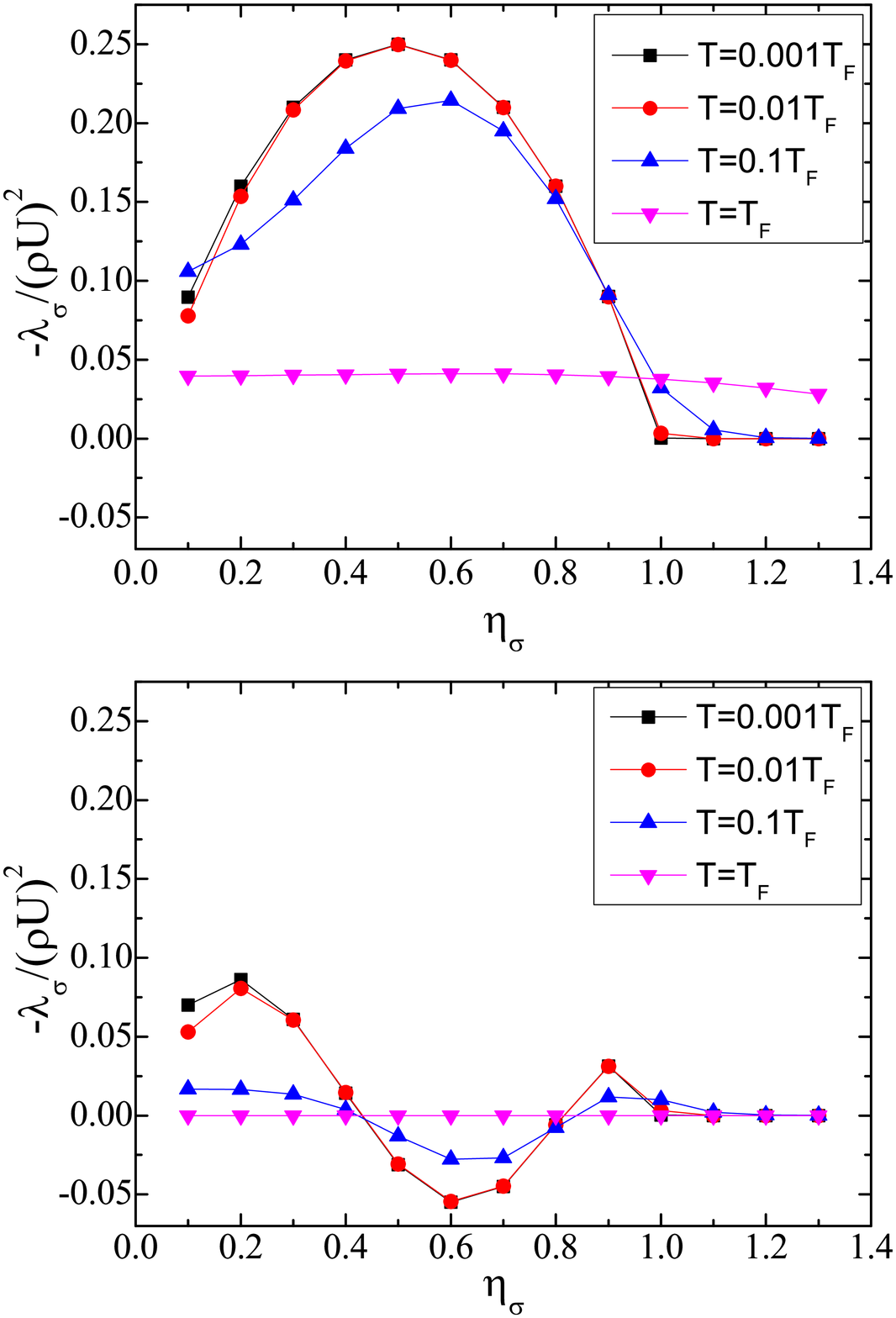}
  \caption{(Color online) Strength of the pairing interactions (
    $-\lambda_{\sigma} /(\rho U)^2$) for spin spieces
    $\sigma$ in the $p$-wave (top) and $f$-wave (bottom)
    channels at different temperatures, where $\eta_{\sigma}$ is the imbalance
    ratio defined as $\eta_{\sigma}=K_{F-\sigma}/K_{F\sigma}$. The scales of the vertical
    axes of these two figures are chosen to be the same for convenient
    comparison.}
  \label{fig:lam}
\end{figure}

In this section, we will determine
the critical temperature by solving the flow equations
numerically. As explained above, at critical temperature, we have
\begin{equation}
  \label{eq:34}
  -\lambda_{\sigma
    m}^{-1}=F_{\sigma}(\Omega_H)-F_{\sigma}(0).
\end{equation}

For convenience, we can define some dimensionless parameters as follows
$x_{\sigma}=k/K_{F \sigma},t_{\sigma}=T/T_{F\sigma}$, where $T_F$ is the
Fermi temperature ($T_F^{\sigma}=\mu_{\sigma}/k_B$), 
and rewrite the susceptibility and the eigenvalues to the final form
for numerical calculation.

Susceptibility only depends on the magnitude of
momentum and can be written as
\begin{eqnarray}
  \label{eq:18}
  \chi_{\sigma}(x_{\sigma}) & =  & \frac{1}{(2\pi)^2}
  \int_0^{\infty}p\mathrm{d}p \int_0^{2\pi}\mathrm{d} \theta
  f(\epsilon_{p,\sigma})
  \frac{2}{\epsilon_{p,\sigma}-\epsilon_{p+k,\sigma}} \nonumber \\
  &=& -\rho \int_0^1 \mathrm{d}y 
  \frac{y}{\sqrt{1-y^2}} \frac{1}{e^{((x_{\sigma}y/2)^2-1)/t_{\sigma}}+1}.
\end{eqnarray}

Eigenvalues of the dimensionless coupling function $g_{\sigma}$ have
the form
\begin{eqnarray}
  \label{eq:20}
  \lambda_{\sigma m} & = &\rho U^2
  \int_0^{2\pi}\frac{\mathrm{d}\theta}{2\pi}
  \chi_{-\sigma}(2k_{F\sigma}\sin \frac{\theta}{2})\cos(m\theta).
\end{eqnarray}

For the $p$-wave and $f$-wave case, we have respectively
\begin{eqnarray}
  \label{eq:35}
  \lambda_{\sigma 1} & = & -(\rho U)^2 \frac{2}{\pi}
  \int_0^1 \mathrm{d}x  \int_0^1 \mathrm{d}y 
  \frac{y}{\sqrt{1-y^2}}\frac{1-2x^2}{\sqrt{1-x^2}} \nonumber   \\
  && \times \frac{1}{e^{(x^2y^2-(K_{F-\sigma}/K_{F\sigma})^2)/t_{\sigma}}+1},
\end{eqnarray}
\begin{eqnarray}
  \label{eq:36}
  \lambda_{\sigma 3} & = & -(\rho U)^2 \frac{2}{\pi}
  \int_0^1 \mathrm{d}x  \int_0^1 \mathrm{d}y 
  \frac{y}{\sqrt{1-y^2}} \nonumber \\
  && \times \frac{-32x^6+48x^4-18x^2+1}{\sqrt{1-x^2}} \nonumber   \\
  && \times \frac{1}{e^{(x^2y^2-(K_{F-\sigma}/K_{F\sigma})^2)/t_{\sigma}}+1}
\end{eqnarray}

\begin{figure}[t]
\centerline{\includegraphics[width=3in]{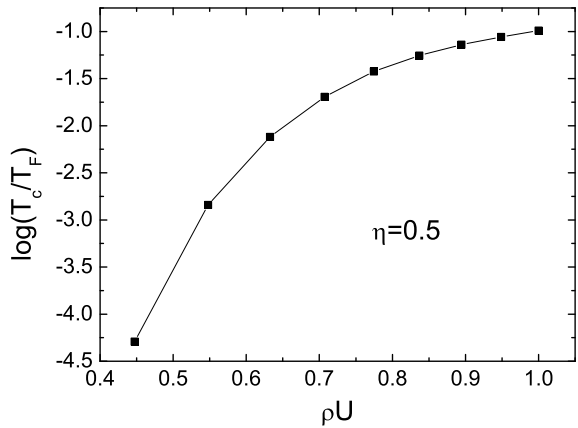}}
\caption[]{\label{fig:tc-u} Critical temperature ($T_c$) from weak coupling
  regime to intermediate coupling regime with imbalanced ratio
  ($\eta$) set to 
$0.5$  near the optimal value. $T_F$ is the Fermi temperature. The
variation of $T_c$ with respect to $\eta$ at different coupling
strengths will be illustrated in Fig.~\ref{fig:tc}.}
\end{figure}

First, the eigenvalues of $p$-wave ($\lambda_{\sigma 1}$) and $f$-wave
($\lambda_{\sigma 3}$) channel are obtained at
different imbalance ratios ($\eta_{\sigma}=K_{F-\sigma}/K_{F\sigma}$)
and temperatures as illustrated in
Fig.~\ref{fig:lam}, where it can be seen that eigenvalues of the $p$-wave
channel is more negative than those in $f$-wave channel, indicating the
leading instability in $p$-wave channel. Eigenvalues are
almost zero when $\eta_{\sigma}>1$, which means that there is no obvious instability for
smaller Fermi surface (we will focus on the majority speices in the
remaining part of this paper). This is similar to the $A_1$ phase of $^3\mathrm{He}$
\cite{Vollhardt1990}  
when applied a magnetic field which will cause spin
population imbalance. According to the qualitative arguments on $A_1$ phase given
by Leggett \cite{Leggett1975},
the reason why pairing only happens for the bigger Fermi surface is
that the density of states at the bigger Fermi surface is larger, which
results in higher critical temperature for the majority species.
However, in a 2D system,
density of states is a constant and we are thus facing a different
situation from the $A_1$ phase of $^3\mathrm{He}$. Besides, we can see that, for
$p$-wave channel, the optimal imbalance ratio where the most negative
eigenvalue appears is about $0.5$, which is consistent with Ref.~\cite{Raghu2011}.
When temperature goes higher and becomes comparable with the Fermi
temperature, the 
eigenvalues are concealed under thermal fluctuation.

To guarantee the validation of the perturbative approach in the first
stage, the dimensionless coupling $\rho U$ should be small. The
induced vertex, which is of order $U^2$, will be even smaller, and
exponentially suppress the critical temperature. Considering
the large-N emerging in the second stage of RG \cite{Shankar1994}, we
can 
extrapolate the results to the intermediate coupling
regime \cite{Raghu2011}.

Setting imbalance ratio to $0.5$, near the optimal value, we plotted
the critical temperature from weak to intermediate coupling regime
(see Fig.~\ref{fig:tc-u}). It can be seen that the critical temperature
drops quickly as the interaction strength goes smaller.

\begin{figure}[t]
  \centering
  \includegraphics[width=3in]{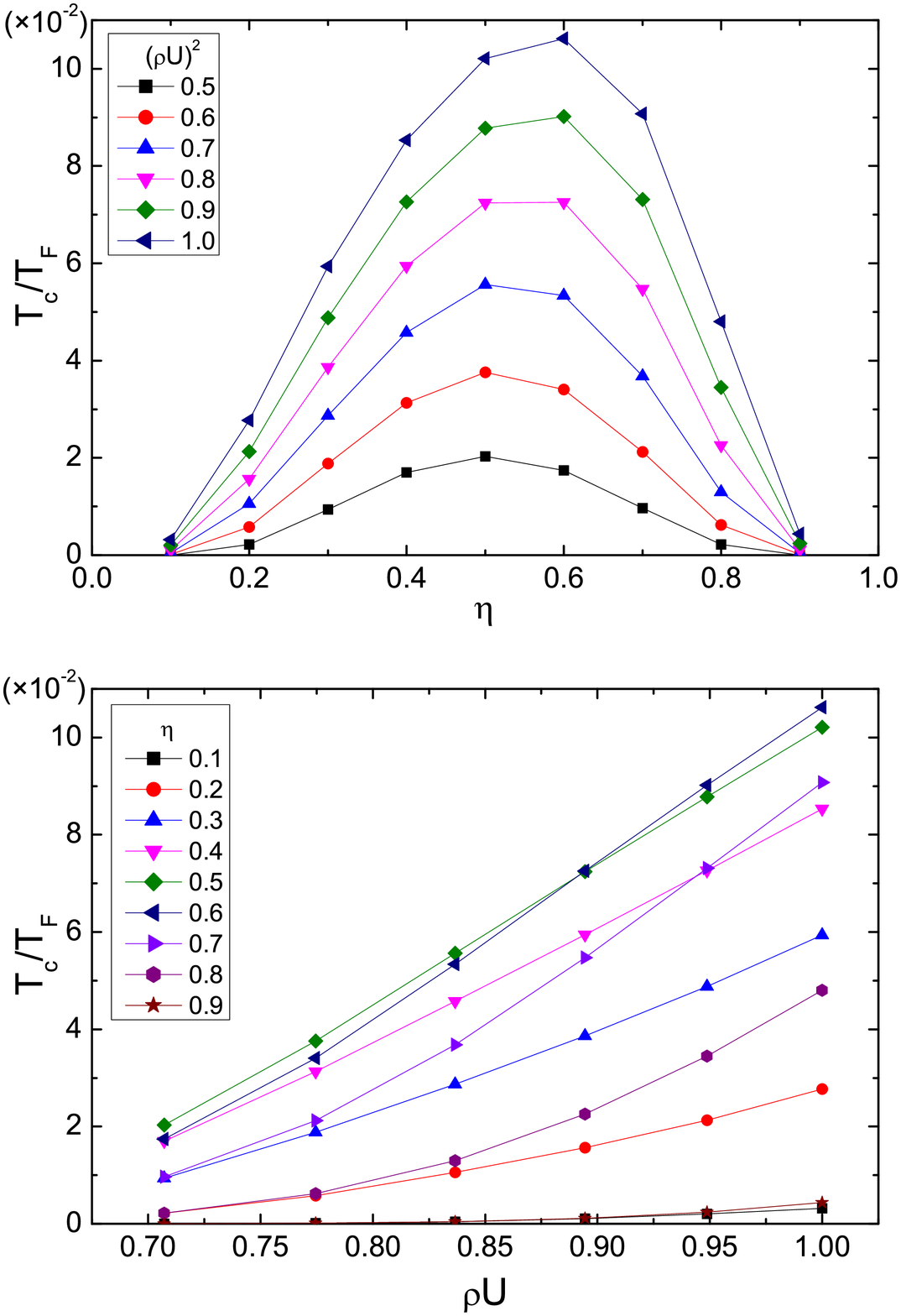}
  \caption{(Color online) Critical temperatures $T_c$ in unit of $T_F$ at different
    imbalance ratios ($\eta$) and
    coupling strengths ($\rho U$) in the intermediate coupling
    regime. The variation of $T_c$ with respect to $\rho U$ from the
    weak to the intermediate coupling regime is shown in Fig.~\ref{fig:tc-u}.}
  \label{fig:tc}
\end{figure}

We also calculated critical temperatures at different imbalance
ratios with 
coupling strengths $\rho U\sim 1$ in the intermediate regime (see
Fig.~\ref{fig:tc}).  For fixed coupling, the highest critical temperature
appears near $\eta=0.5$ as expected and is about $10^{-2}T_F$ but
also drops quickly as the imbalance ratio tends to zero or unity.

\section{Discussions and Summaries}

We have shown that the population imbalance induced $p$-wave
superfluid state may be observable in a 2D repulsive fermion
gas. Population imbalance can be achieved by an unequal mixing of
atoms in two hyperfine states, and tunable repulsive interactions can
be realized by using the upper branch of a Feshbach resonance. In
Ref.\cite{Jo2009}, $^6\mathrm{Li}$ atoms in the repulsive regime were used to
study the itinerant ferromagnetism. One problem that should be
considered 
is that the upper branch of a Feshbach resonance is an excited branch,
and will decay to the BEC molecule state due to inelastic
three-body collisions \cite{Petrov2003}. However, with small scattering length and
population imbalance, the decay rate is suppressed \cite{Jo2009,DIncao2005} and the system may
be metastable for observation.
For experimental observations, we suggest to look for rotational
asymmetries 
in the momentum distribution or pairwise correlation in the time of
flight expansion images of the dominant species
\cite{Altman2004,Greiner2005}.
In addition, the transition 
temperature can be raised in two different ways (see Fig.~\ref{fig:tc}). One is to adjust the imbalance
ratio to the optimal value which is around $0.5$. As can be seen from
Fig.~\ref{fig:tc-u}, the theoretical transition 
temperature should be 
around $10^{-5}T_F \sim 10^{-3}T_F$ in the weak repulsive
regime near the optimal imbalance ratio. Another one is to
increase the coupling strength by Feshbach 
resonance. By extending our result to the intermediate coupling
regime, we get an estimation of the critical temperature which 
reaches as high as
 $10^{-2}T_F$. Since our approach is 
asymptotically exact, the perturbative calculations are well controlled
in the first stage of RG. After safely arriving at the second stage,
where the cutoff $\Lambda$ is much smaller than the Fermi momentum
$K_F$, the emergence of a large-N ensures the non-perturbative nature of
the momentum shell RG in the second stage where a quantitatively
calculation of critical temperature was given. However, for comparing with the
experimental results, we should notice  that another important issue
is the trap effects on our system. In striking contrast with $s$-wave
superfluid state, the trap asymmetries would have a strong influence
on the spontaneously preferred orientation of $p$-wave superfluid
state. We will study the trap effects in our future work.

In summary, we studied a possible new superfluid state for a 2D
population imbalanced fermion gas with short-range 
repulsive interactions. This phenomena is different from the ones in the
BEC-BCS crossover 
where the BEC  
molecule state is concerned. It is also different from  the ones in
the  unitary 
region 
where the scattering length is approaching infinity and many universal
properties emerge. For the system considered in this paper, the bare interaction
is purely repulsive, and there are no intermediate bosons for inducing
attractive interactions.
 We studied this system based on RG
approach and treated different instabilities on an 
equal footing. There are no 
assumptions of specific orders compared with mean-field approach.
What is essential for the Cooper instability is the mismatch of the
Fermi surfaces caused by population imbalance in our system. It can
also be achieved in mixture of fermions with unequal mass, to which
our approach can be generalized straightforwardly.
 By working in the finite temperature formulism and
numerically solving the flow equation, we gave a quantitatively
calculation of the critical temperature.
Our study is of particular significance both for probing $p$-wave
superfluidity in the novel regime experimentally and for studying
imbalanced fermionic systems with RG theory theoretically.

\begin{acknowledgments}
   We acknowledge insightful comments by F. Zhou, Q. Zhou and helpful
  discussions with R. Qi. This work was supported by NSFC under grants
  Nos. 10934010, 60978019, the NKBRSFC under grants
  Nos. 2009CB930701, 2010CB922904, 2011CB921502, 2012CB821300, and
  NSFC-RGC under grants Nos. 11061160490 and 1386-N-HKU748/10.
\end{acknowledgments}

\bibliography{imb.bib}

\end{document}